\documentclass{emulateapj}

\shorttitle{Origin of $^{48}$Ca}

\shortauthors{Wanajo et al.}

\begin{document}

\title{Electron-capture supernovae as origin of $^{48}$C\lowercase{a}}

\author{Shinya Wanajo\altaffilmark{1},
        Hans-Thomas Janka\altaffilmark{2},
        and 
        Bernhard M\"uller\altaffilmark{2}
        }

\altaffiltext{1}{National Astronomical Observatory of Japan,
        2-21-1 Osawa, Mitaka, Tokyo 181-8588, Japan;
        shinya.wanajo@nao.ac.jp}

\altaffiltext{2}{Max-Planck-Institut f\"ur Astrophysik,
        Karl-Schwarzschild-Str. 1, D-85748 Garching, Germany}

\begin{abstract}
We report that electron-capture supernovae (ECSNe), arising from
 collapsing oxygen-neon-magnesium cores, are a possible source of
 $^{48}$Ca, whose origin has remained a long-standing puzzle. Our
 two-dimensional, self-consistent explosion model of an ECSN predicts
 ejection of neutron-rich matter with electron fractions $Y_\mathrm{e}
 \approx 0.40$--0.42 and relatively low entropies, $s \approx 13$--$15\,
 k_\mathrm{B}$ per nucleon ($k_\mathrm{B}$ is the Boltzmann
 constant). Post-processing nucleosynthesis calculations result in
 appreciable production of $^{48}$Ca in such neutron-rich and
 low-entropy matter during the quasi-nuclear equilibrium and subsequent
 freezeout phases. The amount of ejected $^{48}$Ca can account for that
 in the solar inventory when we consider possible uncertainties in the
 entropies. ECSNe could thus be a site of $^{48}$Ca production in
 addition to a hypothetical, rare class of high-density Type~Ia
 supernovae.
\end{abstract}


\section{Introduction}

The origin of the most neutron-rich (n-rich) stable isotope among the
iron-group nuclei and lighter elements, $^{48}$Ca, remains a
long-standing mystery of nucleosynthesis \citep[see][for the historical
background]{Hartmann1985, Meyer1996}.  $^{48}$Ca constitutes only
0.187\% of the solar calcium abundance \citep{Asplund2009}, but is 47
times more abundant than its next heavier stable isotope, $^{46}$Ca.
From the point of view of nuclear stability, this can be attributed to
the doubly magic structure of $^{48}$Ca with $Z = 20$ and $N = 28$.
Thermal equilibrium, such as nuclear statistical equilibrium (NSE) or
quasi-nuclear equilibrium (QSE), thus seems to be responsible for the
creation of $^{48}$Ca, in which each isotopic abundance is a strong
function of its binding energy per nucleon.  Such a formation of
$^{48}$Ca demands a highly n-rich environment with $Y_\mathrm{e}$
(number of protons per nucleon) being close to that characterizing the
structure of $^{48}$Ca, $Y_\mathrm{e, nuc} = 0.417$.

\citet{Hartmann1985} suggested n-rich NSE as the production mechanism of
$^{48}$Ca, based on the assumption that such conditions would be met in
the innermost ejecta of core-collapse supernovae (CCSNe).
\citet{Meyer1996} argued, however, that QSE was responsible for the
creation of $^{48}$Ca. They showed that a low-entropy environment,
characterized by $\phi < 1$, led to the appreciable creation of
$^{48}$Ca, where $\phi \equiv 0.34\, T_9^3/\rho_5$ is the
photon-to-nucleon ratio with $T_9$ being the temperature in $10^9$~K and
$\rho_5$ the matter density in $10^5$~g~cm$^{-3}$. For this reason,
\citet{Meyer1996} concluded that CCSNe were excluded as the origin of
$^{48}$Ca because of their high entropies ($\phi > 1$) in the
neutrino-driven ejecta. Such a high $\phi$ leads to an $\alpha$-rich
freezeout from NSE \citep{Woosley1992, Meyer1998a} followed by QSE, in
which nuclear stability prefers heavier nuclei than $^{48}$Ca. Instead,
they concluded Type~Ia supernovae (SNe~Ia) as the unique site that could
achieve such low $\phi$ environments.  \citet{Woosley1997} explored
SN~Ia models with a central density of (4--$8) \times 10^9$~g~cm$^{-3}$
(needed to obtain $Y_\mathrm{e} \sim 0.40$--0.42), which was
substantially higher than the typical values of about $2 \times
10^9$~g~cm$^{-3}$.  He found the production of $^{48}$Ca/$^{56}$Fe to be
about 100 times larger than its solar ratio. Given the observational
fact that about one half of the Galactic iron comes from SNe~Ia
\citep{Timmes1995}, \citet{Woosley1997} estimated an event rate of such
high-density explosions no more than a few percent of the observed
SNe~Ia rate.

In this Letter, we propose that electron-capture SNe (ECSNe), a
sub-class of CCSNe arising from collapsing O-Ne-Mg cores, can be
additional sites of $^{48}$Ca production. A recent two-dimensional (2D),
self-consistent simulation of an ECSN predicts the ejection of n-rich
matter of $Y_\mathrm{e} \sim 0.40$--0.42 with $s \sim 14\,
k_\mathrm{B}/\mathrm{nuc}$ \citep[$k_\mathrm{B}$ is the Boltzmann
constant;][]{Janka2008, Janka2012, Wanajo2011}. These entropies, or
$\phi \sim 1.4$, are only slightly above the limit of the criterion for
$^{48}$Ca formation considered by \citet{Meyer1996}.  We utilize this
ECSN model (\S~2) for nucleosynthesis calculations and discuss the
production mechanism of $^{48}$Ca (\S~3). The mass-integrated
nucleosynthetic yields are then compared with the solar abundances to
test if our ECSN model accounts for the $^{48}$Ca abundance in the solar
inventory (\S~4).

\section{ECSN model}\label{sec:model}

\begin{figure}
\epsscale{1.0}
\plotone{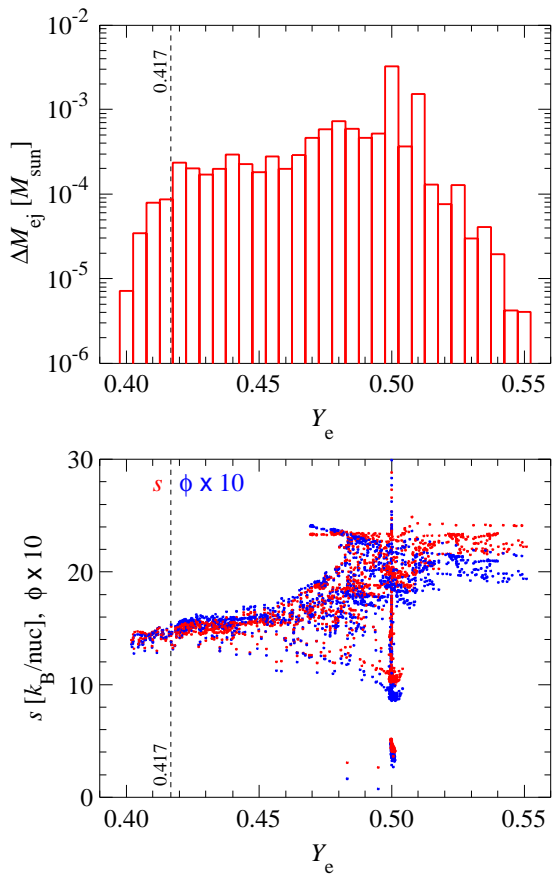}
\caption{Top: Ejecta masses vs. $Y_\mathrm{e}$ at $T_9 = 5$ for our ECSN
 model. The width of the $Y_\mathrm{e}$-bins is chosen to
 be $\Delta Y_\mathrm{e} = 0.005$. 
 The total ejecta mass from the core is $1.14 \times 10^{-2}
 M_\odot$. 
 Also indicated
 by a dashed line is $Y_\mathrm{e, nuc} = 0.417$.
 Bottom: $s$ and $\phi$ 
 as functions of $Y_\mathrm{e}$ at $T_9 = 5$ for all tracer
 particles.
}
\label{fig:yedms}
\end{figure}

The nucleosynthesis analysis makes use of about 2000 representative
tracer particles, by which the thermodynamic histories of ejecta chunks
were followed in our self-consistent 2D hydrodynamic calculation of an
ECSN \citep{Janka2008, Janka2012, Wanajo2011}. The model was computed
with a ray-by-ray-plus treatment of the energy-dependent neutrino
transport \citep{Rampp2000, Buras2006, Kitaura2006}.  The pre-collapse
model of the O-Ne-Mg core emerged from the evolution of an $8.8 M_\odot$
star \citep{Nomoto1987}. In our model, a neutrino-powered explosion sets
in at about 100~ms after core bounce with an energy of about
$10^{50}$~erg \citep{Janka2008}. Because of the very steep density
gradient near the core surface, an interesting amount of n-rich matter
with relatively low entropy ($s\approx 13$--$15\, k_\mathrm{B}$ for
$Y_\mathrm{e} < 0.42$) gets ejected (Figure~\ref{fig:yedms};
top)\footnote{$Y_\mathrm{e}$, $s$, and $\phi$ are those evaluated at
$T_9 = 5$ throughout this paper.}. This happens only in the
multi-dimensional case, which allows for the rapid expansion of
neutrino-heated gas in buoyant, mushroom-like plumes, whose expansion is
too rapid for $\nu_e$ and $\bar\nu_e$ absorptions to lift $Y_\mathrm{e}$
to values closer to 0.5, in contrast to the 1D case or more massive
Fe-core progenitors, where the ejecta expansion is much slower and
neutrino absorption proceeds longer.  The entropies are lower for the
lower $Y_\mathrm{e}$ particles as a result of less neutrino energy
deposition. The photon-to-nucleon ratios (Fig.~\ref{fig:yedms}; bottom)
are approximately one-tenth of the entropies, showing however a slightly
steeper gradient for $Y_\mathrm{e} < 0.42$. The expansion timescales of
the particles, defined as the $e$-folding time of the temperature drop
below 0.5~MeV, are $\tau_\mathrm{exp} = 50$--100~ms and are largely
independent of $Y_\mathrm{e}$.

\section{$^{48}$C\lowercase{a} production}\label{sec:ca48}

\begin{figure}
\epsscale{1.0}
\plotone{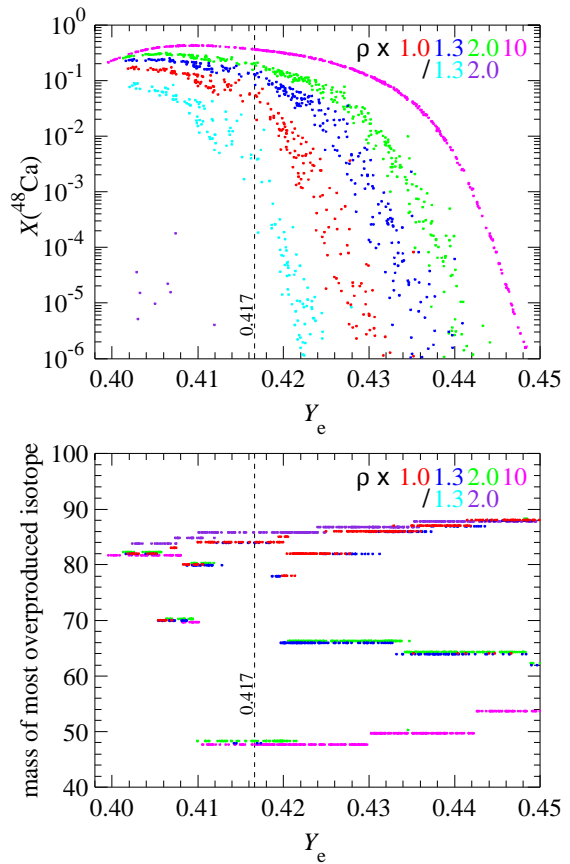}
\caption{Top: Final mass fractions of $^{48}$Ca for the
 tracer particles in the range of $Y_\mathrm{e} < 0.45$.
 Also indicated
 by a dashed line is $Y_\mathrm{e, nuc} = 0.417$.
 The result of the unchanged model is shown in red, and
 those with the densities multiplied by factors of 1.3, 2.0, 10 and
 divided by factors of 1.3, 2.0 are given in different colors. Bottom:
 Mass numbers of nuclei that have the greatest production factors
 ($X/X_\odot$) as functions of $Y_\mathrm{e}$ (slightly shifted in the
 vertical direction for visibility).
}
\label{fig:yeca48}
\end{figure}

\begin{figure*}
\epsscale{1.0}
\plotone{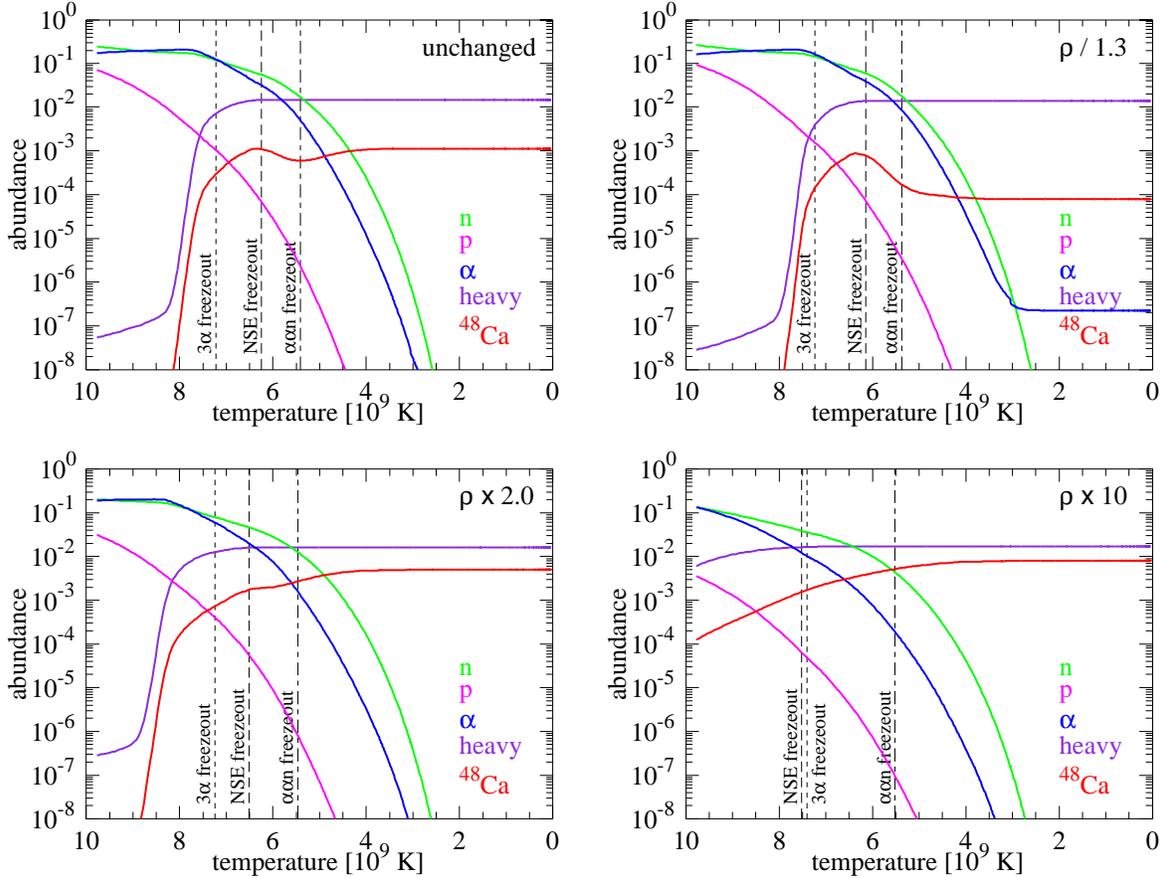}
\caption{Abundances of n, p, $\alpha$, heavy nuclei ($Z > 2$), and
 $^{48}$Ca as functions of descending temperature for the tracer
 particle with $Y_\mathrm{e} = 0.417$, $s =
 14.3\, k_\mathrm{B}/\mathrm{nuc}$ ($\phi = 1.45$), and $\tau_\mathrm{exp} =
 56.7$~ms.  
 The result of the unchanged model is shown in the upper-left, and
 those for the densities divided by a factor of 1.3 and
 multiplied by factors of 2.0, 10 are shown in upper-right, lower-left,
 and lower-right panels, respectively.
 The long-dashed lines mark the NSE-freezeout temperatures.
 Also indicated by dashed and long-short-dashed lines are the
 temperatures at which the $\alpha$-consumption timescales
 evaluated by $3\alpha$ and $\alpha\alpha n$, respectively, become longer than 
 $\tau_\mathrm{exp}$.
}
\label{fig:tabun}
\end{figure*}

\begin{figure}
\epsscale{1.0}
\plotone{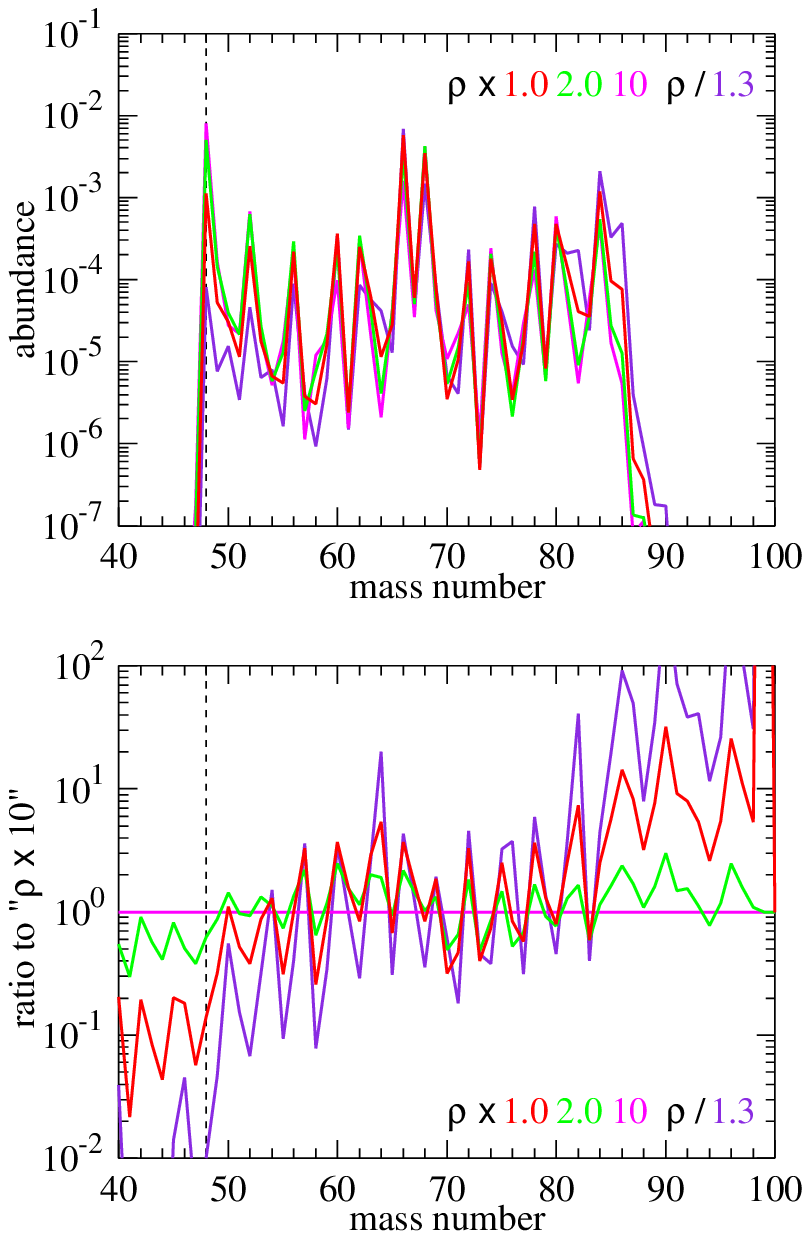}
\caption{Final abundances as functions of mass number (upper panel) for
 the same tracer particle in Fig.~\ref{fig:tabun}.
 The result of the unchanged model is shown in red, and
 those for the densities divided by a factor of 1.3 or
 multiplied by factors of 2.0, 10 are shown in different colors.
 The position of $^{48}$Ca is indicated by a dashed line. The abundance
 ratios relative to the ``$\rho \times 10$'' case are also shown in the
 lower panel.
}
\label{fig:sabun}
\end{figure}

The nucleosynthetic yields are obtained with the reaction network code
described in \citet{Wanajo2009, Wanajo2011} by employing the latest
reaction library of REACLIB V2.0 \citep{Cyburt2010}. Using thermodynamic
trajectories directly from the ECSN model, the calculations are started
when the temperature decreases to $T_9 = 10$, assuming initially free
protons and neutrons with mass fractions $Y_\mathrm{e, i}$ and $1 -
Y_\mathrm{e, i}$, respectively ($Y_\mathrm{e, i}$ is the value at $T_9 =
10$).

The final mass fractions of $^{48}$Ca for the tracer particles in the
range of $Y_\mathrm{e} < 0.45$ are shown as a function of $Y_\mathrm{e}$
by red dots in the upper-panel of Figure~\ref{fig:yeca48}. We find
appreciable production of $^{48}$Ca in the range below its $Y_\mathrm{e,
nuc}$ (dashed line), peaking at $Y_\mathrm{e} \approx
0.40$\footnote{Test calculations with $Y_\mathrm{e, min}$ replaced by
even lower values result in the drop-off of $X(^{48}\mathrm{Ca})$ around
$Y_\mathrm{e} \lesssim 0.390$.}. $^{48}$Ca is always made as itself, its
double-magic nature affording a special degree of stability in nuclear
equilibrium.

To examine the nucleosynthesis of $^{48}$Ca in detail, let us pick a
representative tracer particle with $Y_\mathrm{e} = 0.417$, $s = 14.3\,
k_\mathrm{B}/\mathrm{nuc}$ ($\phi = 1.45$), and $\tau_\mathrm{exp} =
56.7$~ms. Note that the $Y_\mathrm{e}$ is slightly greater than
$Y_\mathrm{e, i} = 0.414$ as a result of neutrino reactions and the
$\alpha$-effect \citep{McLaughlin1996, Meyer1998b}. The abundances
(defined by $Y \equiv X/A$; number per nucleon) of n, p, $\alpha$, heavy
nuclei ($Z > 2$), and $^{48}$Ca are displayed as a function of
descending temperature in panel of Figure~\ref{fig:tabun}
(upper-left). We find that the abundance of heavy nuclei is fixed around
$T_9 = 6.24$. This is a consequence of the fact that the 3-body
reactions $\alpha(\alpha\alpha, \gamma)^{12}$C (hereafter $3\alpha$) and
$\alpha(\alpha n, \gamma)^9$Be followed by $^9$Be$(\alpha, n)$C$^{12}$
(hereafter $\alpha\alpha n$) become slow as the matter expands. The
$\alpha$-consumption timescale, $\tau_{3\alpha} \equiv
-Y_\alpha/(dY_\alpha/dt)$ evaluated by $3\alpha$, first exceeds
$\tau_\mathrm{exp}$ at $T_9 = 7.22$ (dashed line). The gateway from
light to heavy nuclei is closed when the timescale
$\tau_{\alpha\alpha\mathrm{n}}$ evaluated by $\alpha\alpha n$ as well
exceeds $\tau_\mathrm{exp}$ ($T_9 = 5.41$; long-short-dashed line). In
general, NSE freezes between the $3\alpha$ and $\alpha\alpha n$
freezeouts. We define the NSE-freezeout temperature ($T_9 = 6.24$;
long-dashed line) at which the timescale of heavy abundance formation,
$\tau_\mathrm{heavy} \equiv Y_\mathrm{heavy}/(dY_\mathrm{heavy}/dt)$,
exceeds $\tau_\mathrm{exp}$. Once the NSE freezes, the number of heavy
nuclei in QSE cannot decrease because of their net photodisintegration
being too slow \citep{Meyer1996}. The fixed number of heavy nuclei is in
fact the key condition of QSE.

We find that the transition from NSE to QSE takes place with the number
of $\alpha$'s slightly greater than that of heavy nuclei. The abundances
become redistributed by absorptions of light nuclei,\footnote{In QSE,
individual reactions, which are in general much faster than the
expansion timescales of matter, are irrelevant. What determine the
abundance distribution are nuclear binding energies per
nucleon. Individual reactions play roles only after a freezeout from QSE
\citep[$T_9 \approx 4$; see \S~4 in][]{Meyer1998a}. Note also that
$\alpha$'s are in nuclear equilibrium with free nucleons throughout the
QSE phase. All n, p, $\alpha$ thus play roles for readjustment of the
QSE abundances.} decreasing the abundance of $^{48}$Ca and increasing
the abundances of heavier nuclei in the QSE cluster
(Figure~\ref{fig:sabun}). In the late QSE phase ($T_9 \approx 5.5$--4),
however, the $^{48}$Ca abundance recovers in response to a drop-off of
the $\alpha$ abundance. The number of charged light nuclei becomes
insufficient to transmute $^{48}$Ca into heavier nuclei. The abundance
redistribution is predominantly due to photodisintegration and
n-absorption, which cannot make a net upward shift in $Z$. Instead,
$^{48}$Ca increases in the local equilibrium around its double-magic
intersection, lasting until the freezeout from QSE at $T_9 \approx
4$. In the end, however, the most abundant isotope is $^{66}$Zn, the
daughter of p-magic $^{66}$Ni (Fig.~\ref{fig:sabun}).

How can we keep $^{48}$Ca increasing in the QSE phase?  Obviously, the
presence of $\alpha$'s is the cause of shifting light QSE abundances
toward a greater $Z$ in the above case. We thus test the sensitivity of
the $^{48}$Ca production to entropy, the key quantity that controls the
$\alpha$ abundance. Figure~\ref{fig:tabun} gives the results in which
densities are divided by 1.3 (upper-right) or multiplied by 2.0
(lower-left), 10 (lower-right) throughout the calculations for the same
tracer particle. The $\phi$'s (and roughly $s$'s) decrease or increase
by the inverse of the density scaling factor in each case. We find that
the ``$\rho / 1.3$'' case ($\phi = 1.89$) leads to NSE freezeout with
appreciably abundant $\alpha$'s compared to heavy nuclei (hereafter
$\alpha$-rich QSE; $Y_\alpha/Y_\mathrm{heavy} > 1$)\footnote{This is
equivalent to a ``QSE with too few heavy nuclei'' compared to what would
be expected in NSE with the same density and temperature described in
\citet{Meyer1996, Meyer1998a}.}. In this condition, light nuclei are so
abundant that the $^{48}$Ca abundance continuously decreases. In
contrast, for the ``$\rho \times 2$'' case ($\phi = 0.725$), the NSE
freezes with a smaller $\alpha$ abundance that is comparable to that of
heavy nuclei (hereafter $\alpha$-poor QSE; $Y_\alpha/Y_\mathrm{heavy}
\lesssim 1$)\footnote{This is equivalent to a ``QSE with too many heavy
nuclei'' compared to NSE \citep{Meyer1996, Meyer1998a}, in which QSE
cannot reduce the number of heavy nuclei and thus $^{48}$Ca
survives.}. Such low levels of charged light nuclei allow $^{48}$Ca to
continuously increase during QSE and even after the QSE freezeout by
$T_9 \approx 2$. In fact, it is the most abundant isotope in the end
(Fig.~\ref{fig:sabun}). The ``$\rho \times 10$'' case ($\phi = 0.145$)
gives a limiting condition of $\alpha$-poor QSE, resulting in a robust
$^{48}$Ca creation. Figure~\ref{fig:sabun} displays the final abundances
for the same cases and those relative the ``$\rho \times 10$''
case. From the lower panel, it is clearly seen that a more $\alpha$-rich
QSE tends to shift more abundances from lighter to heavier mass
numbers. We conclude, therefore, that $\alpha$-poor QSE is the mechanism
for making $^{48}$Ca as suggested by \citet{Meyer1996}.

Figure~\ref{fig:yeca48} (top) shows the result of sensitivity tests with
various density scaling factors for all the tracer particles. Overall,
the synthesis of $^{48}$Ca is quite sensitive to entropy (or $\phi$) in
the range $\phi \gtrsim 1$ as shown in \citet{Meyer1996}. In the
explored range, $^{48}$Ca is more robustly produced in the more n-rich
environment, $Y_\mathrm{e} < Y_\mathrm{e, nuc} = 0.417$. As an example,
the particles with $Y_\mathrm{e} \approx 0.402$ in the ``$\rho / 1.3$''
case have similar $^{48}$Ca abundances to those of the unchanged case
with $Y_\mathrm{e} \approx 0.417$, despite larger $\phi$'s for the
former ($\approx 1.77$) than the latter ($\approx 1.44$). This is a
consequence of the $\alpha\alpha n$ reaction more rapid in more n-rich
matter \citep{Meyer1998a}, leading to a faster consumption of
$\alpha$'s. It is important to note, however, that $^{48}$Ca is nowhere
the most overproduced isotope (i.e., with the greatest $X/X_\odot$) in
the explored range of our unchanged model as shown in
Figure~\ref{fig:yeca48} (bottom). For $Y_\mathrm{e} < 0.417$, $^{82}$Se
or $^{84}$Kr are generally the most overproduced isotopes. Only the
``$\rho \times 2.0$'' and ``$\rho \times 10$'' cases, corresponding to
$\phi \lesssim 0.7$, lead to the production of $^{48}$Ca as the most
overproduced isotope over a certain range in $Y_\mathrm{e}$. In summary,
the physical conditions of our ECSN model lie in the transition region
between the $\alpha$-rich ($\phi \gtrsim 1$) and $\alpha$-poor ($\phi
\lesssim 1$) QSEs \citep[see Fig.~1 in][]{Meyer1996}.

\section{Contribution to the Galaxy}\label{sec:galaxy}

\begin{figure*}
\epsscale{1.0}
\plotone{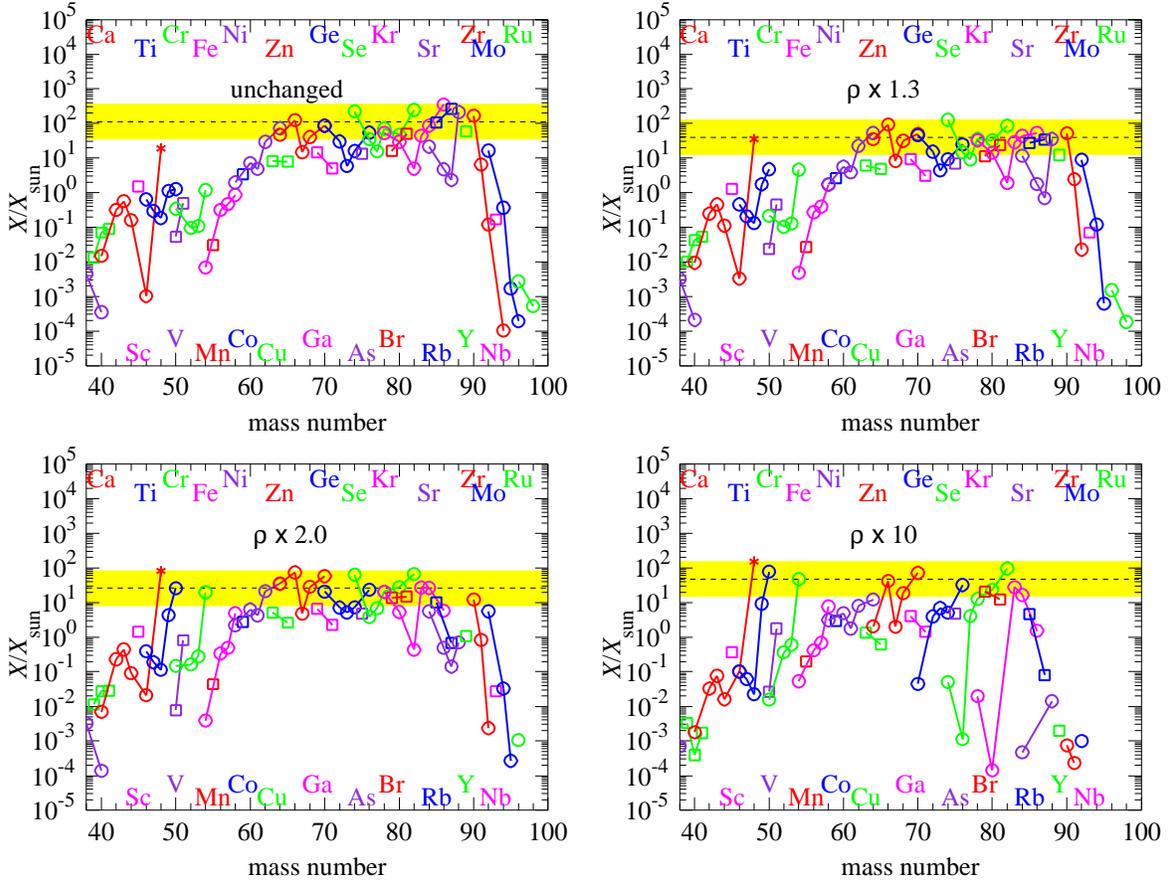}
\caption{Isotopic mass fractions in the ECSN ejecta relative to their
 solar values \citep{Lodders2003} 
 for the unchanged model (upper-left), 
 and the
 models in which the densities are multiplied by 1.3 (upper-right), 2.0
 (lower-left), and 10 (lower-right).
 $^{48}$Ca is indicated by an asterisk in each panel.
 The normalization band (see the text) is marked
 in yellow with the medium value indicated by a dashed line in each panel.
}
\label{fig:pfiso}
\end{figure*}

To examine whether ECSNe can be significant producers of $^{48}$Ca in
the Galaxy, the final abundances are obtained by mass integration over
all the tracer particles. Figure~\ref{fig:pfiso} (upper-left) gives the
isotopic mass fractions (after all radioactivities have decayed)
relative to their solar values \citep{Lodders2003}, i.e., the
``production factors'', as a function of mass number for our unchanged
model.  The ``normalization band'' between the maximum (352 for
$^{86}$Kr) and one-tenth of that is indicated in yellow with the medium
marked by a dashed line. Isotopes that fall into this band can be
accepted as being made in considerable amounts by ECSNe. A reasonable
``flatness'' of production factors can be seen between $A \sim 60$ and
90, in which many isotopes fall into the normalization band. $^{48}$Ca
(shown by asterisk) is however located slightly below the normalization
band. This implies that ECSNe contribute to the production of $^{48}$Ca
no more than 10\% of its solar content.

As explored in \S~\ref{sec:ca48}, the nucleosynthesis of $^{48}$Ca is
highly sensitive to entropy and our unchanged model appears to lie in
the edge of the $\alpha$-poor QSE condition. As a sensitivity test to
entropy, the mass-integrated production factors for ``$\rho \times
1.3$'', ``$\rho \times 2.0$'', and ``$\rho \times 10$'' are also
displayed in the upper-right, lower-left, and lower-right panels of
Figure~\ref{fig:pfiso}, respectively. The result of ``$\rho \times
1.3$'' satisfies the condition of $^{48}$Ca being in the middle of the
normalization band, which means that the model can accounts for
one-third of $^{48}$Ca in the solar system. In order to fully account
for the origin of $^{48}$Ca, one needs to increase the densities (that
is equivalent to reduce $\phi$'s) by a factor of two
(Fig.~\ref{fig:pfiso}; lower-left). The reduced entropies lead to
$\alpha$-poor QSE conditions in a broad range of $Y_\mathrm{e}$, in
which $^{48}$Ca is copiously produced (Figure~\ref{fig:yeca48}; top). In
fact, it is the most overproduced isotope for $0.41 \lesssim
Y_\mathrm{e} \lesssim 0.42$ in this case (Figure~\ref{fig:yeca48};
bottom). $^{50}$Ti and $^{54}$Cr, the isotopes made in $\alpha$-poor
QSE, are also enhanced. The change here corresponds to a reduction of
the entropies from $s \sim 14\, k_\mathrm{B}/\mathrm{nuc}$ to $\sim 11\,
k_\mathrm{B}/\mathrm{nuc}$ or $\sim 7\, k_\mathrm{B}/\mathrm{nuc}$ in
the range $Y_\mathrm{e} < 0.42$. This does not seem to be an extreme
assumption, although the reduced values tend to scratch the lower end of
the range of entropies that are compatible with neutrino-heated ejecta
of ECSNe or CCSNe. In the limiting case of ``$\rho \times 10$'',
$^{48}$Ca remains the most overproduced isotope, although it is not much
increased from the ``$\rho \times 2.0$'' case. This is a consequence of
robust $^{48}$Ca production for $\phi < 1$ as proved in
\citet{Meyer1996}. It is interesting to note that the result in this
case resembles to that of high-density SN~Ia by
\citet{Woosley1997}. Isotopes made in $\alpha$-rich QSE, such as
$^{64}$Zn, $^{70}$Ge, $^{74}$Se, and $^{78, 86}$Kr are all depleted and
a reasonable flatness of production factors cannot be seen.

The nucleosynthesis of $^{48}$Ca also depends on $\tau_\mathrm{exp}$.
In principle, a slower expansion leads to a later freezeout from NSE
and, consequently, leads to $\alpha$-poor QSE more easily. Such a
consequence, however, is not the case in our ECSN model. More slowly
expanding matter would be exposed to neutrino absorptions for a longer
time, shifting $Y_\mathrm{e}$ toward higher values as a result of the
$\alpha$-effect. Such a $Y_\mathrm{e}$ increase would compensate the
benefits of a slower expansion. One may also consider to increase the
ejecta masses where $^{48}$Ca is richly produced ($Y_\mathrm{e} \approx
0.40$--0.42). In fact, fine details of the $Y_\mathrm{e}$-$\Delta
M_\mathrm{ej}$ distribution could well depend on the spatial resolution
in the simulation or on other factors. It is not plausible, however, to
enhance $^{48}$Ca without others in our unchanged model. As found in the
lower panel of Figure~\ref{fig:yeca48}, the most overproduced isotope in
the range $Y_\mathrm{e} < 0.42$ is generally $^{82}$Se or $^{84}$Kr that
already lie on the normalization band (Fig.~\ref{fig:pfiso};
upper-left).

\section{Implications}

We studied the nucleosynthesis of $^{48}$Ca using the thermodynamic
trajectories of a self-consistent 2D explosion model of an ECSN
\citep{Janka2008, Wanajo2011}. Appreciable production of $^{48}$Ca was
found in $\alpha$-poor QSE conditions. Comparisons of the
mass-integrated yields with the solar abundances show underproduction of
$^{48}$Ca. This problem would be cured if the entropies in the ejecta
were somewhat lower than those in our original model. We conclude,
therefore, that ECSNe can be, at least in part, the astrophysical source
of $^{48}$Ca. The fact that ECSNe can be also associated with the origin
of many other isotopes \citep[of light trans-iron species and potentially
of weak r-process species,][]{Wanajo2011} seems encouraging.

Our result also has implications for the interpretation of anomalies of
$^{48}$Ca found in meteorites \citep{Lee1978, Moynier2010, Chen2011}. So
far, only hypothetical high-density SNe~Ia have been considered as the
cause of these anomalies. ECSNe could be, however, an additional source
of the anomalies of $^{48}$Ca. ECSNe may occur at about one-tenth of the
rate of normal CCSNe \citep{Ishimaru1999, Wanajo2011}, that is, 1--2
events per millennium, which makes them much more common than the
rare-class SNe~Ia \citep[once in 10,000 years,][]{Woosley1997}. This
would be important if the composition of the proto-solar system had been
affected by a single or a few nearby SNe.

More work is needed before final conclusions about the role of ECSNe as
production sites of $^{48}$Ca can be drawn.  In particular, further
improvements of ECSN models (e.g., 3D, general relativity, resolution)
will be important to elucidate whether appreciable amounts of n-rich
ejecta of ECSNe satisfy the $\alpha$-poor QSE condition needed for the
nucleosynthesis of $^{48}$Ca.

\acknowledgements

S.W. was supported by the JSPS Grants-in-Aid for Scientific Research
(23224004). At Garching, support by Deutsche Forschungsgemeinschaft
through grants SFB/TR7 and EXC-153 is acknowledged.

\end{document}